\documentclass[twocolumn,showpacs,preprintnumbers,amsmath,amssymb,prl,superscriptaddress]{revtex4}
\usepackage{graphicx}
\usepackage{dcolumn}
\usepackage{bm}
\usepackage{txfonts}
\usepackage{amsmath}
\usepackage{amsfonts}

\begin{document}

\title{Direct measurement of a pure spin current by a polarized light beam\footnote{This is a corrected version of
Phys. Rev. Lett. {\bf 100}, 086603 (2008), where the interaband transition Hamiltonian in Eq.~(5) contains errors in
the spin notations which leads to errors in the effective coupling Hamiltonian in Eqs.~(7) and (10).
Unchanged is the main conclusion that a pure spin current can be detected by a polarized light beam
through a current-current coupling, but the Voigt effect now is absent and the Faraday rotation is enhanced and
has different dependence on the incident angles.}}
\author{Jing Wang}
\affiliation{Department of Physics, The Chinese University of Hong Kong, Shatin, N.T., Hong Kong, China}
\affiliation{Department of Physics, Tsinghua University, Beijing 100084, China}
\author{Bang-Fen Zhu}
\affiliation{Department of Physics, Tsinghua University, Beijing 100084, China}
\affiliation{Center for Advanced Study, Tsinghua University, Beijing 100084, China}
\author{Ren-Bao Liu} \thanks{To whom correspondence should be addressed: rbliu@phy.cuhk.edu.hk}
\affiliation{Department of Physics, The Chinese University of Hong Kong, Shatin, N.T., Hong Kong, China}
\date{\today}

\begin{abstract}
The photon helicity  may be mapped to a spin-1/2, whereby we put
forward an intrinsic interaction between a polarized light beam as a
``photon spin current'' and a pure spin current in a semiconductor,
which arises from the spin-orbit coupling in valence bands as a pure
relativity effect without involving the Rashba or the Dresselhaus
effect due to inversion asymmetries. The interaction leads to
circular optical birefringence, which is similar to the Faraday
rotation in magneto-optics but nevertheless involve no net
magnetization. The birefringence effect provide a direct,
non-demolition measurement of pure spin currents.
\end{abstract}


\maketitle

Measuring spin currents is an indispensable part of spin-based electronics~\cite{Spintronics_Review2}.
Spin-polarized currents flow with net magnetization and therefore can be sensed by conventional
Faraday or Kerr rotation~\cite{Kikkawa:1999,Stephens:2004,Crooker:2005} or through ferromagnetic
filters~\cite{Lou:2007,Appelbaum:2007}. A pure spin current consists of two counter-flowing
currents with equal amplitude but opposite spin polarizations and thus bears neither net
charge current nor net magnetization. Pure spin currents have been inferred from a few
pioneering experiments, being converted either into charge/voltage
signals~\cite{Valenzuela:2006,Ganichev:2007,Cui:2007} via the spin Hall
effect~\cite{Dyakonov:1971,Hirsch:1999,Murakami:2003,Sinova:2004},
or into spin-polarized electrons or excitons where the currents
terminate~\cite{Kato_2004,Wunderlich:2005,Stevens:2003,Zhao:2006}.
An open and fundamental question is: How can a pure spin current be directly measured?

\begin{figure}[t]
\begin{center}
\includegraphics[width=\columnwidth]{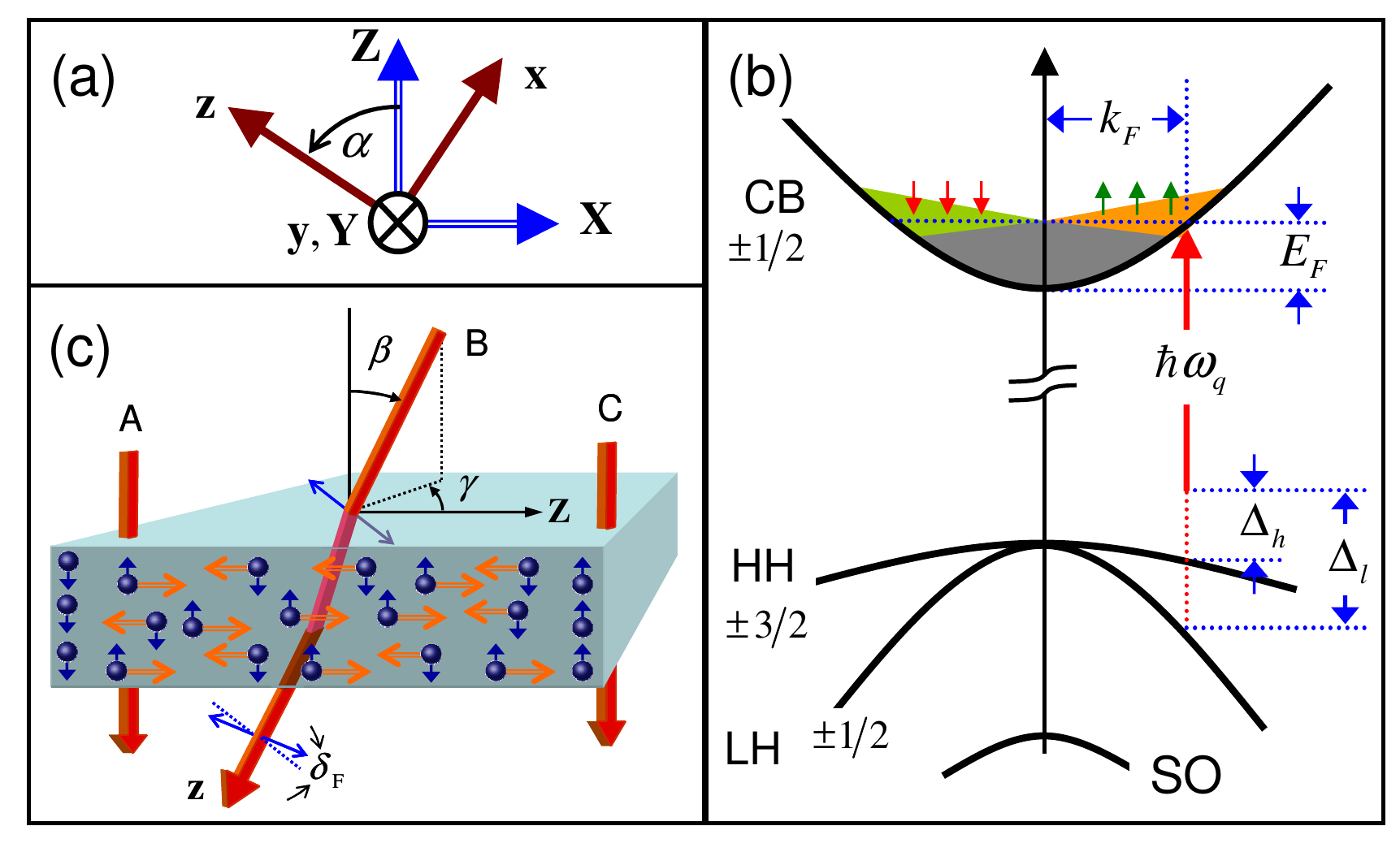}
\end{center}
\caption{ (color online) (a) The spin current direction (${\mathbf
Z}$), the light beam direction (${\mathbf z}$), and the associated
coordinate systems. Without loss of generality, ${\mathbf y}$ and
${\mathbf Y}$ are set parallel. (b) Schematic band structure near
the $\Gamma$ point of an n-doped III-V compound semiconductor, and
illustration of a non-equilibrium electron distribution for a pure
spin current. (c) Setup for measuring a transverse pure spin
current. Case A and C show the measurement of spins accumulated at
the edges by the Faraday rotation of a normal incident light as in
Ref.~\cite{Kato_2004}. Case B shows the direct measurement of the
spin current in the middle region by the Faraday rotation of an
oblique light beam.} \label{geometry}
\end{figure}

A clue to the answer is the Amp\`ere effect where a charge current, even with net charge density
being neutral everywhere, can be detected by other moving charges or another current.
A straightforward analogue suggests that a pure spin current could be measured
by another spin current. In general, for a current breaking the intrinsic symmetry of a system,
a phenomenological coupling can be constructed with another current of the same symmetry-breaking
type. A pure spin current conserves the time-inversion symmetry but breaks the rotation and
space-inversion symmetries, and can be formulated as a rank-2 pseudo-tensor
\begin{equation}
{\mathbb J}=J_X{\mathbf X}{\mathbf Z}+J_Y{\mathbf Y}{\mathbf Z}+J_Z{\mathbf Z}{\mathbf Z}\equiv {\mathbf J}{\mathbf Z},
\end{equation}
where ${\mathbf Z}$ is the current direction, ${\mathbf X}$ and ${\mathbf Y}$ are the transverse
directions [see Fig.~\ref{geometry}~(a)], and ${\mathbf J}$ is along the spin polarization direction.
To form a phenomenological current-current coupling
in a system of all the fundamental symmetries, the probe current should be of the same tensor type.
In lieu of a real spin current for the probe, we would rather use a polarized light beam which
is more feasible. A light can be regarded as a ``photon spin current''~\cite{Note_on_pure_current}
by mapping the photon polarization into a spin-1/2 with the Jones vector representation~\cite{Jones_vector},
$\cos\frac{\theta}{2}e^{i\phi/2}{\mathbf n}_+ +\sin\frac{\theta}{2}e^{-i\phi/2}{\mathbf n}_-\sim \left|\theta,\phi\right\rangle$,
where the right/left circular polarization ${\mathbf n}_{+/-}$ corresponds to the spin up/down
state $\left|\uparrow/\downarrow\right\rangle$ quantized along the light propagation direction.
The ``spin current'' tensor for a light with electric field
$
{\mathbf F}({\mathbf r},t) =
     \left(F_{+}{\mathbf n}_{+}+F_{-}{\mathbf n}_{-}\right) e^{i{\mathbf q}\cdot{\mathbf r}-i\omega_{q}t}    +{\rm c.c.}
$
can be formulated as
\begin{subequations}
\begin{eqnarray}
{\mathbb I} &\equiv & q\left(I_x{\mathbf x}{\mathbf z}+I_y{\mathbf y}{\mathbf z}+I_z{\mathbf z}{\mathbf z}\right)\equiv q {\mathbf I}{\mathbf z}, \\
I_j & = & \frac{1}{2} \sum_{\mu,\nu=\pm}\sigma^j_{\mu\nu}F^*_{\mu}F_{\nu},
\end{eqnarray}
\end{subequations}
where $\sigma^j$ ($j=x, y, z$) is the Pauli matrix, and the unit
axis vectors ${\mathbf x}$, ${\mathbf y}$, and ${\mathbf z}$ are
defined by ${\mathbf n}_{\pm}\equiv \left(\mp{\mathbf x} - i{\mathbf
y}\right)/\sqrt{2}$ and ${\mathbf z}\equiv {\mathbf q}/q$.

The specific form of the phenomenological coupling depends on the microscopic
mechanisms. Consider an n-doped bulk III-V compound semiconductor with a direct
band gap (such as GaAs) and a pure spin current due to a steady non-equilibrium
distribution $\hat{\rho}$ [see Fig.~\ref{geometry} (b)]. An effective coupling
between a polarized beam and the spin current can be mediated by the interband
virtual excitations. Since the light polarization essentially couples only to
the orbital motion of electrons, spin-orbit interaction is required to establish
the effective coupling. As there is inherent spin-orbit coupling in the valence
bands as a relativity effect, the Rashba or Dresselhaus effect due to inversion
asymmetries~\cite{Rashba,Dresselhaus,Rashba_Dresselhaus_splitting} is not a necessity.

Assuming the light is tuned below the Fermi surface and near the
band edge, we consider optical transitions between the conduction
band (CB) and the heavy-hole (HH) and light-hole (LH) bands, and
neglect the split-off band (SO) [see Fig.~\ref{geometry}~(b)].
The Luttinger-Kohn Hamiltonian $h_{\rm LK}$ for the valence bands
near the band edge is~\cite{HandbookOnSemiconductors}
\begin{equation}\label{H_L}
h_{\rm LK} = \frac{\hbar^2}{2m}\left[\left(\gamma_1 +\frac{5}{2}\gamma_2\right){\boldsymbol \nabla}^2
  - 2\gamma_2\left({\boldsymbol \nabla} \cdot {\mathbf K}\right)^2\right]
\end{equation}
where ${\mathbf K}$ is a spin-3/2 for the total angular momentum of a hole.
We have for simplicity neglected the valence-band anisotropy which would not change
the essential results in this paper. The energy dispersion for a hole with the magnetic
quantum number $K_{\mathbf p}$ quantized along the momentum ${\mathbf p}$ is
$E_{K_{\mathbf p}}({\mathbf p}) = \left[\gamma_1+\left(5/2-2K_{\mathbf p}^2\right)\gamma_2\right]\hbar^2p^2/\left({2m}\right)$.
The non-interacting Hamiltonian for electrons and holes is
\begin{equation}
\hat{H}_0 =
\sum_{\pm,{\mathbf p}}\left(
 E_{e{\mathbf p}}\hat{e}_{\pm{\mathbf p}}^{\dagger}\hat{e}_{\pm{\mathbf p}}
+E_{h{\mathbf p}}\hat{h}_{\pm{\mathbf p}}^{\dagger}\hat{h}_{\pm{\mathbf p}}
+E_{l{\mathbf p}}\hat{l}_{\pm{\mathbf p}}^{\dagger}\hat{l}_{\pm{\mathbf p}}
\right),
\end{equation}
where $\hat{e}_{\pm{\mathbf p}}$ annihilates an electron with spin $\pm 1/2$,
$\hat{h}_{\pm{\mathbf p}}$ and $\hat{l}_{\pm,{\mathbf p}}$
annihilate heavy and light holes with $K_{\mathbf p}=\pm 3/2$ and $\pm 1/2$,
respectively, $E_{h{\mathbf p}} =E_{\pm{3}/{2}}({\mathbf p})=\hbar^2 p^2/(2m_h)$,
$E_{l{\mathbf p}}=E_{\pm1/2}({\mathbf p})=\hbar^2p^2/(2m_l)$, and $E_{e{\mathbf p}}=\hbar^2p^2/(2m_e)$.
Here $m_e$, $m_h$, and $m_l$ are in turn the electron, HH, and LH effective mass.
Notice that the angular momentum ${\mathbf K}$ is quantized along ${\mathbf p}$ so that
the spin-orbit coupling in the valence bands is automatically included.
There is no Rashba effect in the bulk system, and the Dresselhaus spin splitting
is negligible~\cite{Note_on_Dresselhaus}.

The interband transition has the Hamiltonian 
\begin{eqnarray}
\hat{H}_1&=&
{d_{\rm cv}^*}\sum_{\mu,\nu,{\mathbf p}}
  F^*_{\nu}{\mathbf n}_{\nu}^*\cdot
  \Big({\mathbf n}_{\mu,{\mathbf p}}\hat{h}_{\mu-{\mathbf p}}\hat{e}_{\bar{\mu}{\mathbf q}+{\mathbf p}}+
\nonumber \\ && \frac{1}{\sqrt{3}}{\mathbf n}_{\mu,{\mathbf
p}}\hat{l}_{\mu-{\mathbf p}}\hat{e}_{\mu{\mathbf q}+{\mathbf p}}
-\sqrt{\frac{2}{3}}{\mathbf z}_{\mathbf p}\hat{l}_{\mu-{\mathbf
p}}\hat{e}_{\bar{\mu}{\mathbf q}+{\mathbf p}}\Big)+{\rm h.c.}, \ \ \
\
\end{eqnarray}
with ${\mathbf n}_{\pm,{\mathbf p}}$ denoting the right/left
circular polarization about ${\mathbf p}$ which are defined as
$\mathbf{n}_{\pm,\mathbf{p}}\equiv\left(\mp\mathbf{x}_{\mathbf{p}}-i\mathbf{y}_{\mathbf{p}}\right)/\sqrt{2}$,
${\mathbf z}_{\mathbf p}\equiv {\mathbf p}/p$, and $\bar{\mu}\equiv
-{\mu}$.

The effective Hamiltonian between the light beam and the spin current is obtained by
the second-order perturbation as
\begin{equation}
{\mathcal H}_{\rm eff} = {\rm Tr}\left[\hat{\rho} \hat{H}_1 \left(\hbar\omega_q-\hat{H}_0\right)^{-1} \hat{H}_1\right],
\end{equation}
and explicitly worked out to be
\begin{subequations}
\begin{eqnarray}
{\mathcal H}_{\rm eff}&=& -{|d_{\mathrm{cv}}|^2}
\sum_{\sigma,\sigma^{\prime}}F^*_{\sigma}F_{\sigma^{\prime}}{{\mathbf
n}_{\sigma^{\prime}}{\mathbf n}^*_{\sigma}}: \label{line1}
 \\
&&\sum_{\mu,{\mathbf p}}\left[f_{\bar{\mu}_{\mathbf
p}\bar{\mu}_{\mathbf p}, {\mathbf q}+{\mathbf p}}
        \frac{{\mathbf n}_{\mu,{\mathbf p}}{\mathbf n}^*_{\mu,{\mathbf p}}}
         {E_{e{\mathbf q}+{\mathbf p}}+E_{h-{\mathbf p}}-\hbar\omega_q}\right.
\label{line2}
 \\
&&+\frac{1}{3}f_{{\bar{\mu}}_{\mathbf p}{\bar{\mu}}_{\mathbf p},
{\mathbf q}+{\mathbf p}}
        \frac{{\mathbf n}_{\bar{\mu},{\mathbf p}}{\mathbf n}^*_{\bar{\mu},{\mathbf p}}+2{\mathbf z}_{\mathbf p}{\mathbf z}_{\mathbf p}^*}
         {E_{e{\mathbf q}+{\mathbf p}}+E_{l-{\mathbf p}}-\hbar\omega_q}
\label{line3}
 \\
&&\left. -\frac{\sqrt{2}}{3}f_{\mu_{\mathbf p}\bar{\mu}_{\mathbf p},
{\mathbf q}+{\mathbf p}}
                    \frac{{\mathbf n}_{\bar{\mu},{\mathbf p}}{\mathbf z}_{\mathbf p}^*+{\mathbf z}_{\mathbf p}{\mathbf n}_{{\mu},{\mathbf p}}^*}
                     {E_{e{\mathbf q}+{\mathbf p}}+E_{l-{\mathbf p}}-\hbar\omega_q}\right],
\label{line4}
\end{eqnarray}
\label{detail_effective}
\end{subequations}
where $f_{\mu\nu,{\mathbf p}}\equiv {\rm
Tr}\left[\hat{\rho}\hat{e}^{\dag}_{\mu{\mathbf
p}}\hat{e}_{\nu{\mathbf p}}\right]$, and $\mu_{\mathbf p}$ indicates
that the spin is quantized along ${\mathbf p}$. Here we have omitted
the trivial background constant. The physical processes for
different terms are identified as follow: (\ref{line2}) accounts for
the HH-CB transitions where a (virtually) absorbed photon has to be
emitted with the same circular polarization and thus the electron
spin is conserved. A LH state contains both circular and linear
orbital components, so a LH-CB transition can be either circularly
or linearly polarized. In (\ref{line3}), the LH-CB transitions in
virtual absorption and emission have the same polarization and in
turn the electron spin is conserved. In (\ref{line4}), the
transitions from and to the LH bands have different polarizations,
leading to orbital angular momentum transfer between the light and
the electrons, while the total angular momentum is still conserved.

As we are not interested in charge effects such as a charge current, we shall drop
spin-independent populations like $f_{++,{\mathbf p}}+f_{--,{\mathbf p}}$ (which
contributes only a change of the background refraction index) but keep only the spin distribution
${\mathbf s}({\mathbf p})\equiv \frac{1}{2}\sum_{\mu,\nu} {\boldsymbol \sigma}_{\mu,\nu}f_{\mu\nu,{\mathbf p}}$.
Furthermore, we consider a pure spin current and assume the net spin polarization
of the system is zero. The spin current
is defined by the dyadic of the spin polarization and the velocity, summed over the momentum space:
\begin{eqnarray}
{\mathbb J}\equiv e\sum_{{\mathbf p}} {\mathbf s}({\mathbf p})
                     \hbar^{-1}{\boldsymbol \nabla}_{\mathbf p}E_{e{\mathbf p}}.
\label{spincurrent}
\end{eqnarray}
A pure spin current results when the electrons of opposite velocities have opposite spins.
With the charge background dropped and the total spin polarization absent, the leading contribution
to the coupling in Eq.~(\ref{detail_effective}) would come from the spin current as discussed below.

To proceed from Eq.~(\ref{detail_effective}), let us first neglect
the small light wavevector ${\mathbf q}$. Since the light excites
the same electron spin for opposite electron momenta $\pm{\mathbf
p}$, the polarized light is only coupled to a net spin
polarization, but not to a pure spin current.

Then we include the small light wavevector ${\mathbf q}$ up to its
first order by the expansion $E_{e{\mathbf q}+{\mathbf p}}\approx
E_{e{\mathbf p}}+{\mathbf q}\cdot{\boldsymbol \nabla}_{\mathbf
p}E_{e{\mathbf p}}$ and $f_{\mu,\nu}({\mathbf q}+{\mathbf p})\approx
f_{\mu,\nu}({\mathbf p}) +{\mathbf q}\cdot{\boldsymbol
\nabla}_{\mathbf p}f_{\mu,\nu}({\mathbf p})$. The gradient in the
momentum space ${\boldsymbol\nabla}_{\mathbf p}$ contributes the
electron velocity which is opposite for opposite momenta. So the light
couples to opposite electron spin for opposite momenta, and
in turn to a pure spin current.

To determine the specific form of the effective Hamiltonian,
we assume that the electron distribution deviates only slightly from the
equilibrium with Fermi wavevector $k_F$ [or Fermi energy $E_F$, see Fig.~\ref{geometry}~(b)]
and the spin distribution has the form
\begin{equation}
{{\mathbf s}}({\mathbf p})={\mathbf s}(p)\cos\theta_{\mathbf p},
\label{distribution}
\end{equation}
where $\theta_{\mathbf p}$ is the angle between ${\mathbf p}$ and
the current direction ${\mathbf Z}$. Such a distribution is usually the case for weak currents.
The light frequency is lower by $\Delta_h$ and $\Delta_l$ than the transition energy
from the HH and the LH bands to the  Fermi surface, respectively.
A straightforward integration over the momentum space yields the effective coupling as
\begin{eqnarray}
{\mathcal H}_{\rm eff} &=& \zeta_1 q {I}_z{\mathbf z}\cdot{\mathbb
J}\cdot {\mathbf z}  + \zeta_2 q I_zJ_Z, \label{effective}
\end{eqnarray}
with
\begin{eqnarray}
\zeta_1 &\equiv& \frac{\hbar{|d_{\mathrm{cv}}|^2}}{e} \left(\frac{4
m_e}{5\Delta^2_h m_h} +\frac{8 m_e}{15\Delta^2_l m_l}
           -\frac{1}{5\Delta_h E_F}+\frac{1}{5\Delta_l E_F}  \right),\nonumber \\
\zeta_2 &\equiv&\frac{\hbar{|d_{\mathrm{cv}}|^2}}{e} \left(\frac{2
m_e}{5\Delta^2_h m_h} -\frac{2 m_e}{5\Delta^2_l m_l}
            -\frac{3}{5\Delta_h E_F}+\frac{3}{5\Delta_l E_F}  \right). \nonumber
\end{eqnarray}
For a spin distribution different from Eq.~(\ref{distribution}), the coupling constants
shown above will only be quantitatively changed in some form factors.
Obviously, the effective coupling, being a tensor contraction between the pure spin
current and the ``photon spin current'', has all the fundamental symmetries of the system.

The linear optical susceptibility is directly related to the phenomenological Hamiltonian by
\begin{equation}
\chi_{\mu,\nu}+\chi^*_{\nu,\mu}=\frac{1}{\epsilon_0}\frac{{\partial^2
{\mathcal H}_{\rm eff}}}{{\partial F^*_{\mu}\partial F_{\nu}}}.
\label{chi}
\end{equation}
Here $\epsilon_0$ is the vacuum permittivity. Since the light is
tuned off-resonant from the available transitions, no real
absorption will occur, but different phaseshifts for different light
polarizations would be induced. In other words, a pure spin current
would produce circular birefringence effect, which is similar to the
Faraday rotation in magnetooptics~\cite{Magnetooptics}, but
nonetheless involves no net magnetization.

The two terms in Eq.~(\ref{effective}) depends on the small light
wavevector $q$, characteristic of coupling to the magnetic dipole
moment of a chiral quantity. Indeed, both $J_Z$ and ${\mathbf
z}\cdot {\mathbb J}\cdot{\mathbf z}$ are longitudinal components of
the spin current in which the current propagation and polarization
are parallel, and hence are chiral. The chiral spin current is
coupled to a circularly polarized photon current, $I_z$, which is
also a chiral quantity. These $q$-dependent terms, stemming from
total angular momentum conserving processes [(\ref{line2}),
(\ref{line3}) and (\ref{line4})], would not cause the circular
polarizations of a light to be flipped but rather induce a
polarization-dependent phaseshift. The susceptibility for opposite
circular polarizations are opposite:
\begin{equation}
\chi_{++}=-\chi_{--}=\frac{1}{4\epsilon_0}q\left(\zeta_1{\mathbf
z}\cdot{\mathbf J}{\mathbf Z}\cdot{\mathbf z}
       + \zeta_2 J_Z\right),
\label{FR}
\end{equation}
which results in circular birefringence similar to the Faraday
rotation~\cite{Magnetooptics} but with no net magnetization. The
Faraday rotation angle is
\begin{equation}
\delta_{\rm F}={\omega_q
L}\left(\chi_{++}-\chi_{--}\right)/\left(4nc\right).
\end{equation}
where $L$ is the light propagation distance, $n$ is the material
refractive index, $c$ is the light velocity in vacuum.

To demonstrate the feasibility of using the optical birefringence
for direct measurement of a pure spin current, we consider a
realistic case that was studied in Ref.~\cite{Kato_2004} where a
transverse spin current is caused by the spin Hall effect. In
Ref.~\cite{Kato_2004}, the Faraday rotation of a light normal to the
sample surface (and parallel to the spin polarization) is measured,
with non-vanishing results only near the edges where spins are
accumulated [case A and C in Fig.~\ref{geometry}~(c)]. The absence
of Faraday rotation in the middle region where the spin current
flows with no net spin polarization is readily explained by
Eq.~(\ref{FR}): In the experimental setup, ${\mathbf Z}\cdot
{\mathbf z}=0$ and $J_Z=0$. To directly measure the spin current
where it flows, we propose to tilt the light beam with a zenith
angle ${\beta}$ from the normal direction and an azimuth angle
$\gamma$ from the spin current direction and detect the Faraday
rotation $\delta_{\rm F}$ in the middle region [Case B in
Fig.~\ref{geometry}~(c)]. The results are predicted to vary with the
incident angles as
\begin{eqnarray}
\delta_{\rm F}(\beta,\gamma)&=&\delta_{\rm F,0}\sin\beta\cos\gamma.
\end{eqnarray}
To estimate the amplitudes of the effect, we use the parameters for
the GaAs sample in Ref.~\cite{Kato_2004}, i.e., $m_e=0.067m_0$
($m_0$ being the free electron mass),
      $m_h=0.45m_0$,
      $m_l=0.082m_0$,
      $L=2.0$~$\mu$m, $n=3.0$, $d_{\mathrm{cv}}=6.7$~$e$\AA, and
      $E_F=5.3$~meV ($k_F=0.96\times 10^6$~cm$^{-1}$ for doping density $3\times 10^{16}$~cm$^{-3}$).
      We take the light wavelength to be around $800$~nm, and the detuning $\Delta_h=1.0$~meV
      and $\Delta_l=4.5$~meV.
For a spin current with amplitude
$J_s=20$~nA$\mu$m$^{-2}$~\cite{Kato_2004}, the maximum Faraday
rotation, reached when $\beta\rightarrow \pi/2$ and
$\gamma\rightarrow 0$, is $\delta_{\rm F,0}\approx 0.38$~$\mu$rad,
measurable in experiments~\cite{Kato_2004}.

In conclusion, the intrinsic interaction between a polarized light
and a spin current may induce measurable circular birefringence as a
direct non-demolition measurement of a pure spin current. Unlike the
generation of spin
currents~\cite{Ganichev:2007,Cui:2007,Bhat:2000,Bhat:2005,Culcer:2007,Ganichev:2003b},
the measurement scheme proposed here does not rely on the inversion
asymmetry of the system.

This work was supported by the Hong Kong RGC Direct Grant No. 2060284,
the NSFC Grant Nos. 10774086, 10574076 and the Basic Research Program of China Grant No. 2006CB921500.
R. B. L. is grateful to X. Dai and L. J. Sham for discussions.


\begin{thebibliography}{30}
\expandafter\ifx\csname natexlab\endcsname\relax\def\natexlab#1{#1}\fi
\expandafter\ifx\csname bibnamefont\endcsname\relax
  \def\bibnamefont#1{#1}\fi
\expandafter\ifx\csname bibfnamefont\endcsname\relax
  \def\bibfnamefont#1{#1}\fi
\expandafter\ifx\csname citenamefont\endcsname\relax
  \def\citenamefont#1{#1}\fi
\expandafter\ifx\csname url\endcsname\relax
  \def\url#1{\texttt{#1}}\fi
\expandafter\ifx\csname urlprefix\endcsname\relax\def\urlprefix{URL }\fi
\providecommand{\bibinfo}[2]{#2}
\providecommand{\eprint}[2][]{\url{#2}}

\bibitem[{\citenamefont{Zutic et~al.}(2004)\citenamefont{Zutic, Fabian, and
  Das~Sarma}}]{Spintronics_Review2}
\bibinfo{author}{\bibfnamefont{I.}~\bibnamefont{Zutic}},
  \bibinfo{author}{\bibfnamefont{J.}~\bibnamefont{Fabian}}, \bibnamefont{and}
  \bibinfo{author}{\bibfnamefont{S.}~\bibnamefont{Das~Sarma}},
  \bibinfo{journal}{Rev. Mod. Phys.} \textbf{\bibinfo{volume}{76}},
  \bibinfo{pages}{323 } (\bibinfo{year}{2004}).

\bibitem[{\citenamefont{Kikkawa and Awschalom}(1999)}]{Kikkawa:1999}
\bibinfo{author}{\bibfnamefont{J.~M.} \bibnamefont{Kikkawa}} \bibnamefont{and}
  \bibinfo{author}{\bibfnamefont{D.~D.} \bibnamefont{Awschalom}},
  \bibinfo{journal}{Nature} \textbf{\bibinfo{volume}{397}}, \bibinfo{pages}{139
  } (\bibinfo{year}{1999}).

\bibitem[{\citenamefont{Stephens et~al.}(2004)\citenamefont{Stephens,
  Berezovsky, McGuire, Sham, Gossard, and Awschalom}}]{Stephens:2004}
\bibinfo{author}{\bibfnamefont{J.}~\bibnamefont{Stephens}},
  \bibinfo{author}{\bibfnamefont{J.}~\bibnamefont{Berezovsky}},
  \bibinfo{author}{\bibfnamefont{J.~P.} \bibnamefont{McGuire}},
  \bibinfo{author}{\bibfnamefont{L.~J.} \bibnamefont{Sham}},
  \bibinfo{author}{\bibfnamefont{A.~C.} \bibnamefont{Gossard}},
  \bibnamefont{and} \bibinfo{author}{\bibfnamefont{D.~D.}
  \bibnamefont{Awschalom}}, \bibinfo{journal}{Phys. Rev. Lett.}
  \textbf{\bibinfo{volume}{93}}, \bibinfo{pages}{097602}
  (\bibinfo{year}{2004}).

\bibitem[{\citenamefont{Crooker et~al.}(2005)\citenamefont{Crooker, Furis, Lou,
  Adelmann, Smith, Palmstr{\o}m, and Crowell}}]{Crooker:2005}
\bibinfo{author}{\bibfnamefont{S.~A.} \bibnamefont{Crooker}},
  \bibinfo{author}{\bibfnamefont{M.}~\bibnamefont{Furis}},
  \bibinfo{author}{\bibfnamefont{X.}~\bibnamefont{Lou}},
  \bibinfo{author}{\bibfnamefont{C.}~\bibnamefont{Adelmann}},
  \bibinfo{author}{\bibfnamefont{D.~L.} \bibnamefont{Smith}},
  \bibinfo{author}{\bibfnamefont{C.~J.} \bibnamefont{Palmstr{\o}m}},
  \bibnamefont{and} \bibinfo{author}{\bibfnamefont{P.~A.}
  \bibnamefont{Crowell}}, \bibinfo{journal}{Science}
  \textbf{\bibinfo{volume}{309}}, \bibinfo{pages}{2191 }
  (\bibinfo{year}{2005}).

\bibitem[{\citenamefont{Lou et~al.}(2007)\citenamefont{Lou, Adelmann, Crooker,
  Garlid, Zhang, Reddy, Flexner, Palmstr{\o}m, and Crowell}}]{Lou:2007}
\bibinfo{author}{\bibfnamefont{X.~H.} \bibnamefont{Lou}},
  \bibinfo{author}{\bibfnamefont{C.}~\bibnamefont{Adelmann}},
  \bibinfo{author}{\bibfnamefont{S.~A.} \bibnamefont{Crooker}},
  \bibinfo{author}{\bibfnamefont{E.~S.} \bibnamefont{Garlid}},
  \bibinfo{author}{\bibfnamefont{J.}~\bibnamefont{Zhang}},
  \bibinfo{author}{\bibfnamefont{K.~S.~M.} \bibnamefont{Reddy}},
  \bibinfo{author}{\bibfnamefont{S.~D.} \bibnamefont{Flexner}},
  \bibinfo{author}{\bibfnamefont{C.~J.} \bibnamefont{Palmstr{\o}m}},
  \bibnamefont{and} \bibinfo{author}{\bibfnamefont{P.~A.}
  \bibnamefont{Crowell}}, \bibinfo{journal}{Nature Phys.}
  \textbf{\bibinfo{volume}{3}}, \bibinfo{pages}{197 } (\bibinfo{year}{2007}).

\bibitem[{\citenamefont{Appelbaum et~al.}(2007)\citenamefont{Appelbaum, Huang,
  and Monsma}}]{Appelbaum:2007}
\bibinfo{author}{\bibfnamefont{I.}~\bibnamefont{Appelbaum}},
  \bibinfo{author}{\bibfnamefont{B.~Q.} \bibnamefont{Huang}}, \bibnamefont{and}
  \bibinfo{author}{\bibfnamefont{D.~J.} \bibnamefont{Monsma}},
  \bibinfo{journal}{Nature} \textbf{\bibinfo{volume}{447}}, \bibinfo{pages}{295
  } (\bibinfo{year}{2007}).

\bibitem[{\citenamefont{Valenzuela and Tinkham}(2006)}]{Valenzuela:2006}
\bibinfo{author}{\bibfnamefont{S.~O.} \bibnamefont{Valenzuela}}
  \bibnamefont{and} \bibinfo{author}{\bibfnamefont{M.}~\bibnamefont{Tinkham}},
  \bibinfo{journal}{Nature} \textbf{\bibinfo{volume}{442}}, \bibinfo{pages}{176
  } (\bibinfo{year}{2006}).

\bibitem[{\citenamefont{Ganichev et~al.}(2007)\citenamefont{Ganichev, Danilov,
  Bel'kov, Giglberger, Tarasenko, Ivchenko, Weiss, Jantsch, Sch\"affler, Gruber
  et~al.}}]{Ganichev:2007}
\bibinfo{author}{\bibfnamefont{S.~D.} \bibnamefont{Ganichev}},
  \bibinfo{author}{\bibfnamefont{S.~N.} \bibnamefont{Danilov}},
  \bibinfo{author}{\bibfnamefont{V.~V.} \bibnamefont{Bel'kov}},
  \bibinfo{author}{\bibfnamefont{S.}~\bibnamefont{Giglberger}},
  \bibinfo{author}{\bibfnamefont{S.~A.} \bibnamefont{Tarasenko}},
  \bibinfo{author}{\bibfnamefont{E.~L.} \bibnamefont{Ivchenko}},
  \bibinfo{author}{\bibfnamefont{D.}~\bibnamefont{Weiss}},
  \bibinfo{author}{\bibfnamefont{W.}~\bibnamefont{Jantsch}},
  \bibinfo{author}{\bibfnamefont{F.}~\bibnamefont{Sch\"affler}},
  \bibinfo{author}{\bibfnamefont{D.}~\bibnamefont{Gruber}},
  \bibnamefont{et~al.}, \bibinfo{journal}{Phys. Rev. B}
  \textbf{\bibinfo{volume}{75}}, \bibinfo{pages}{155317}
  (\bibinfo{year}{2007}).

\bibitem[{\citenamefont{Cui et~al.}(2007)\citenamefont{Cui, Shen, Li, Ji, Ge,
  and Zhang}}]{Cui:2007}
\bibinfo{author}{\bibfnamefont{X.~D.} \bibnamefont{Cui}},
  \bibinfo{author}{\bibfnamefont{S.~Q.} \bibnamefont{Shen}},
  \bibinfo{author}{\bibfnamefont{J.}~\bibnamefont{Li}},
  \bibinfo{author}{\bibfnamefont{Y.}~\bibnamefont{Ji}},
  \bibinfo{author}{\bibfnamefont{W.}~\bibnamefont{Ge}}, \bibnamefont{and}
  \bibinfo{author}{\bibfnamefont{F.~C.} \bibnamefont{Zhang}},
  \bibinfo{journal}{Appl. Phys. Lett.} \textbf{\bibinfo{volume}{90}},
  \bibinfo{pages}{242115} (\bibinfo{year}{2007}).

\bibitem[{\citenamefont{Dyakonov and Perel}(1971)}]{Dyakonov:1971}
\bibinfo{author}{\bibfnamefont{M.~I.} \bibnamefont{Dyakonov}} \bibnamefont{and}
  \bibinfo{author}{\bibfnamefont{V.~I.} \bibnamefont{Perel}},
  \bibinfo{journal}{Phys. Lett. A} \textbf{\bibinfo{volume}{35}},
  \bibinfo{pages}{459} (\bibinfo{year}{1971}).

\bibitem[{\citenamefont{Hirsch}(1999)}]{Hirsch:1999}
\bibinfo{author}{\bibfnamefont{J.~E.} \bibnamefont{Hirsch}},
  \bibinfo{journal}{Phys. Rev. Lett.} \textbf{\bibinfo{volume}{83}},
  \bibinfo{pages}{1834 } (\bibinfo{year}{1999}).

\bibitem[{\citenamefont{Murakami et~al.}(2003)\citenamefont{Murakami, Nagaosa,
  and Zhang}}]{Murakami:2003}
\bibinfo{author}{\bibfnamefont{S.}~\bibnamefont{Murakami}},
  \bibinfo{author}{\bibfnamefont{N.}~\bibnamefont{Nagaosa}}, \bibnamefont{and}
  \bibinfo{author}{\bibfnamefont{S.~C.} \bibnamefont{Zhang}},
  \bibinfo{journal}{Science} \textbf{\bibinfo{volume}{301}},
  \bibinfo{pages}{1348 } (\bibinfo{year}{2003}).

\bibitem[{\citenamefont{Sinova et~al.}(2004)\citenamefont{Sinova, Culcer, Niu,
  Sinitsyn, Jungwirth, and Macdonald}}]{Sinova:2004}
\bibinfo{author}{\bibfnamefont{J.}~\bibnamefont{Sinova}},
  \bibinfo{author}{\bibfnamefont{D.}~\bibnamefont{Culcer}},
  \bibinfo{author}{\bibfnamefont{Q.}~\bibnamefont{Niu}},
  \bibinfo{author}{\bibfnamefont{N.~A.} \bibnamefont{Sinitsyn}},
  \bibinfo{author}{\bibfnamefont{T.}~\bibnamefont{Jungwirth}},
  \bibnamefont{and} \bibinfo{author}{\bibfnamefont{A.~H.}
  \bibnamefont{Macdonald}}, \bibinfo{journal}{Phys. Rev. Lett.}
  \textbf{\bibinfo{volume}{92}}, \bibinfo{pages}{126603}
  (\bibinfo{year}{2004}).

\bibitem[{\citenamefont{Kato et~al.}(2004)\citenamefont{Kato, Myers, Gossard,
  and Awschalom}}]{Kato_2004}
\bibinfo{author}{\bibfnamefont{Y.~K.} \bibnamefont{Kato}},
  \bibinfo{author}{\bibfnamefont{R.~C.} \bibnamefont{Myers}},
  \bibinfo{author}{\bibfnamefont{A.~C.} \bibnamefont{Gossard}},
  \bibnamefont{and} \bibinfo{author}{\bibfnamefont{D.~D.}
  \bibnamefont{Awschalom}}, \bibinfo{journal}{Science}
  \textbf{\bibinfo{volume}{306}}, \bibinfo{pages}{1910 }
  (\bibinfo{year}{2004}).

\bibitem[{\citenamefont{Wunderlich et~al.}(2005)\citenamefont{Wunderlich,
  Kaestner, Sinova, and Jungwirth}}]{Wunderlich:2005}
\bibinfo{author}{\bibfnamefont{J.}~\bibnamefont{Wunderlich}},
  \bibinfo{author}{\bibfnamefont{B.}~\bibnamefont{Kaestner}},
  \bibinfo{author}{\bibfnamefont{J.}~\bibnamefont{Sinova}}, \bibnamefont{and}
  \bibinfo{author}{\bibfnamefont{T.}~\bibnamefont{Jungwirth}},
  \bibinfo{journal}{Phys. Rev. Lett.} \textbf{\bibinfo{volume}{94}},
  \bibinfo{pages}{047204} (\bibinfo{year}{2005}).

\bibitem[{\citenamefont{Stevens et~al.}(2003)\citenamefont{Stevens, Smirl,
  Bhat, Najmaie, Sipe, and van Driel}}]{Stevens:2003}
\bibinfo{author}{\bibfnamefont{M.~J.} \bibnamefont{Stevens}},
  \bibinfo{author}{\bibfnamefont{A.~L.} \bibnamefont{Smirl}},
  \bibinfo{author}{\bibfnamefont{R.~D.~R.} \bibnamefont{Bhat}},
  \bibinfo{author}{\bibfnamefont{A.}~\bibnamefont{Najmaie}},
  \bibinfo{author}{\bibfnamefont{J.~E.} \bibnamefont{Sipe}}, \bibnamefont{and}
  \bibinfo{author}{\bibfnamefont{H.~M.} \bibnamefont{van Driel}},
  \bibinfo{journal}{Phys. Rev. Lett.} \textbf{\bibinfo{volume}{90}},
  \bibinfo{pages}{136603} (\bibinfo{year}{2003}).

\bibitem[{\citenamefont{Zhao et~al.}(2006)\citenamefont{Zhao, Loren, van Driel,
  and Smirl}}]{Zhao:2006}
\bibinfo{author}{\bibfnamefont{H.}~\bibnamefont{Zhao}},
  \bibinfo{author}{\bibfnamefont{E.~J.} \bibnamefont{Loren}},
  \bibinfo{author}{\bibfnamefont{H.~M.} \bibnamefont{van Driel}},
  \bibnamefont{and} \bibinfo{author}{\bibfnamefont{A.~L.} \bibnamefont{Smirl}},
  \bibinfo{journal}{Phys. Rev. Lett.} \textbf{\bibinfo{volume}{96}},
  \bibinfo{pages}{246601} (\bibinfo{year}{2006}).

\bibitem[{Not({\natexlab{a}})}]{Note_on_pure_current}
\bibinfo{note}{Strictly speaking, a light beam does not conserve the
  time-inversion symmetry. But its angular momentum flux apart from the energy
  flux, being a pure spin current of interest here, does.}

\bibitem[{\citenamefont{Jones}(1941)}]{Jones_vector}
\bibinfo{author}{\bibfnamefont{R.~C.} \bibnamefont{Jones}},
  \bibinfo{journal}{J. Opt. Soc. Am.} \textbf{\bibinfo{volume}{31}},
  \bibinfo{pages}{488 } (\bibinfo{year}{1941}).

\bibitem[{\citenamefont{Bychkov and Rashba}(1984)}]{Rashba}
\bibinfo{author}{\bibfnamefont{Y.~A.} \bibnamefont{Bychkov}} \bibnamefont{and}
  \bibinfo{author}{\bibfnamefont{E.~I.} \bibnamefont{Rashba}},
  \bibinfo{journal}{Pis'ma Zh. Eksp. Teor. Fiz.} \textbf{\bibinfo{volume}{39}},
  \bibinfo{pages}{66} (\bibinfo{year}{1984}), \bibinfo{note}{[Sov. Phys. JETP
  Lett. {\bf 39}, 78 (1984)]}.

\bibitem[{\citenamefont{Dresselhaus}(1955)}]{Dresselhaus}
\bibinfo{author}{\bibfnamefont{G.}~\bibnamefont{Dresselhaus}},
  \bibinfo{journal}{Phys. Rev.} \textbf{\bibinfo{volume}{100}},
  \bibinfo{pages}{580} (\bibinfo{year}{1955}).

\bibitem[{\citenamefont{Ganichev et~al.}(2004)\citenamefont{Ganichev, Bel'kov,
  Golub, Ivchenko, Schneider, Giglberger, Eroms, and
  De~Boeck}}]{Rashba_Dresselhaus_splitting}
\bibinfo{author}{\bibfnamefont{S.~D.} \bibnamefont{Ganichev}},
  \bibinfo{author}{\bibfnamefont{V.~V.} \bibnamefont{Bel'kov}},
  \bibinfo{author}{\bibfnamefont{L.~E.} \bibnamefont{Golub}},
  \bibinfo{author}{\bibfnamefont{E.~L.} \bibnamefont{Ivchenko}},
  \bibinfo{author}{\bibfnamefont{P.}~\bibnamefont{Schneider}},
  \bibinfo{author}{\bibfnamefont{S.}~\bibnamefont{Giglberger}},
  \bibinfo{author}{\bibfnamefont{J.}~\bibnamefont{Eroms}}, \bibnamefont{and}
  \bibinfo{author}{\bibfnamefont{J.}~\bibnamefont{De~Boeck}},
  \bibinfo{journal}{Phys.\ Rev. Lett.} \textbf{\bibinfo{volume}{92}},
  \bibinfo{pages}{256601} (\bibinfo{year}{2004}).

\bibitem[{\citenamefont{Paul and Moss}(1982)}]{HandbookOnSemiconductors}
\bibinfo{editor}{\bibfnamefont{W.}~\bibnamefont{Paul}} \bibnamefont{and}
  \bibinfo{editor}{\bibfnamefont{T.~S.} \bibnamefont{Moss}}, eds.,
  \emph{\bibinfo{title}{Handbook on Semicondutors: Band Theory and Transport
  Properties}}, vol.~\bibinfo{volume}{1} (\bibinfo{publisher}{North-Holland},
  \bibinfo{address}{Amsterdam}, \bibinfo{year}{1982}).

\bibitem[{Not({\natexlab{b}})}]{Note_on_Dresselhaus}
\bibinfo{note}{The Dresselhaus spliting is $\sim 0.01$~meV in GaAs with doping
  density $\sim 10^{16}$~cm$^{-3}$~\cite{Pikus}, much less than the detuning of
  the light from the Fermi surface we shall use.}

\bibitem[{\citenamefont{Zvezdin and Kotov}(1997)}]{Magnetooptics}
\bibinfo{author}{\bibfnamefont{A.~K.} \bibnamefont{Zvezdin}} \bibnamefont{and}
  \bibinfo{author}{\bibfnamefont{V.~A.} \bibnamefont{Kotov}},
  \emph{\bibinfo{title}{Modern Magnetooptics and Magnetooptical Materials}}
  (\bibinfo{publisher}{Taylor and Francis Group}, \bibinfo{address}{New York},
  \bibinfo{year}{1997}).

\bibitem[{\citenamefont{Bhat and Sipe}(2000)}]{Bhat:2000}
\bibinfo{author}{\bibfnamefont{R.~D.~R.} \bibnamefont{Bhat}} \bibnamefont{and}
  \bibinfo{author}{\bibfnamefont{J.~E.} \bibnamefont{Sipe}},
  \bibinfo{journal}{Phys. Rev. Lett.} \textbf{\bibinfo{volume}{85}},
  \bibinfo{pages}{5432 } (\bibinfo{year}{2000}).

\bibitem[{\citenamefont{Bhat et~al.}(2005)\citenamefont{Bhat, Nastos, Najmaie,
  and Sipe}}]{Bhat:2005}
\bibinfo{author}{\bibfnamefont{R.~D.~R.} \bibnamefont{Bhat}},
  \bibinfo{author}{\bibfnamefont{F.}~\bibnamefont{Nastos}},
  \bibinfo{author}{\bibfnamefont{A.}~\bibnamefont{Najmaie}}, \bibnamefont{and}
  \bibinfo{author}{\bibfnamefont{J.~E.} \bibnamefont{Sipe}},
  \bibinfo{journal}{Phys. Rev. Lett.} \textbf{\bibinfo{volume}{94}},
  \bibinfo{pages}{096603} (\bibinfo{year}{2005}).

\bibitem[{\citenamefont{Culcer and Winkler}(2007)}]{Culcer:2007}
\bibinfo{author}{\bibfnamefont{D.}~\bibnamefont{Culcer}} \bibnamefont{and}
  \bibinfo{author}{\bibfnamefont{R.}~\bibnamefont{Winkler}},
  \bibinfo{journal}{Phys. Rev. Lett.} \textbf{\bibinfo{volume}{99}},
  \bibinfo{pages}{226601} (\bibinfo{year}{2007}).

\bibitem[{\citenamefont{Ganichev and Prettl}(2003)}]{Ganichev:2003b}
\bibinfo{author}{\bibfnamefont{S.~D.} \bibnamefont{Ganichev}} \bibnamefont{and}
  \bibinfo{author}{\bibfnamefont{W.}~\bibnamefont{Prettl}},
  \bibinfo{journal}{J. Phys. - Cond. Matt.} \textbf{\bibinfo{volume}{15}},
  \bibinfo{pages}{R935 } (\bibinfo{year}{2003}).

\bibitem[{\citenamefont{Pikus et~al.}(1988)\citenamefont{Pikus, Marushchak, and
  Titkov}}]{Pikus}
\bibinfo{author}{\bibfnamefont{G.~E.} \bibnamefont{Pikus}},
  \bibinfo{author}{\bibfnamefont{V.~A.} \bibnamefont{Marushchak}},
  \bibnamefont{and} \bibinfo{author}{\bibfnamefont{A.~N.}
  \bibnamefont{Titkov}}, \bibinfo{journal}{Sov. Phys. Semicond.}
  \textbf{\bibinfo{volume}{22}}, \bibinfo{pages}{115} (\bibinfo{year}{1988}).

\end{thebibliography}

\end{document}